# Sampling strategy and statistical analysis for radioactive waste characterization

Nadia Pérot [1*], Alexandre Le Cocguen[1], Dominique Carré[1], Hervé Lamotte[1], Anne Duhart-Barone[1], Ingmar Pointeau[1]

[1]CEA Nuclear Energy Division, Centre de Cadarache, 13108 Saint-Paul-lez-Durance
*Nadia Pérot, nadia.perot@cea.fr

**Abstract**

This paper describes the methodology we have developed to define a sampling strategy adapted to operational constraints in order to characterize the dihydrogen flow rate of 2714 nuclear waste drums produced by radiolysis reaction of organic mixed with $\alpha$-emitters. The objective was to perform few but relevant measurements. Thus, a sample of only 38 drums has been selected to be measured. Statistical analysis of drum measurement data of dihydrogen rate provided an estimation of the mean and the upper bound of the physical quantity of interest which gave a good convergence with global measurements from the ventilation system of the facility. Thereafter, performing a factorial data analysis has demonstrated the representativeness of the measurement data set and the sampling strategy assumption validity. Moreover, it provided information that has been used for a regression analysis to develop a linear prediction model of dihydrogen flow rate production for the waste drum characterization.

## 1. Introduction

Some categories of radioactive waste drums may produce hydrogen gas because of the radiolysis reaction of organic matter like PVC, Polyethylene or cellulose mixed with $\alpha$-emitters in the waste. The evaluation of the hydrogen flow rate produced by radioactive waste drums is required for their disposal in final waste repositories. However, considering the time required for the $H_2$ flow rate measurement of only one drum (more than one month) and the need to characterize the population of 2714 drums (on the CEA center of Cadarache), only a small sample can be measured. Therefore, it is necessary to develop a statistical sampling strategy to select a drum set of "reasonable" size that would be representative of the whole population.

This paper describes the methodology used to define a sampling strategy adapted to the operational constraints of the facility and to analyze the drum measurement data of $H_2$ rate completed by the validation of the sampling strategy hypotheses. The following section presents the sampling strategy performed to identify a representative set of drums to measure and the analysis of the measurement data to estimate statistical indicators of the quantity of interest, the drum $H_2$ flow rate. The third section is dedicated to the validation of the sampling strategy hypothesis and finally, a regression analysis has been performed on the measurement data in order to develop a predictive model of drum $H_2$ production.

## 2. Sampling strategy and statistical analysis

The need to characterize annual $H_2$ production of this drum population corresponds in practice to an objective of upper bound estimation. In this context, the Wilks method has been selected.

## 2.1. Wilks method

The Wilks method has been introduced in the nuclear engineering community by the German nuclear safety institute (GRS) at the beginning of the 1990s (Hoffer 1993), and then used for various safety assessment problems. This method (Wilks 1941) based on order statistics allows the user to precisely determine the required sample size in order to estimate, for a random variable, a quantile[1] of order $\gamma$ with confidence level $\beta$. The great interest of this method (Wilks 1941; Blatman 2017) is its robustness and that no hypothesis is required.

We restrict our explanation below to the one-sided case. Suppose we have an *i.i.d.* (independent identically distributed) n-sample $X_1, X_2, \cdots, X_n$ drawn from a random variable X. We note $M = \max_i(X_i)$. For M to be an upper bound for at least $\gamma.100\%$ of possible values of X with given confidence level $\beta$, we require

$$P[P(X \leq M) \geq \gamma] \geq \beta \quad \text{(Eq. 1)}.$$

The Wilks formula stands that the sample size n must therefore satisfy the following inequality:

$$1 - \gamma^n \geq \beta \quad \text{(Eq. 2)}$$

| $\gamma$ | 0.9 | 0.9 | 0.9 | 0.95 | 0.95 | 0.95 | 0.95 | 0.99 | 0.99 |
|---|---|---|---|---|---|---|---|---|---|
| $\beta$ | 0.5 | 0.9 | 0.95 | 0.4 | 0.5 | 0.78 | 0.95 | 0.95 | 0.99 |
| $n$ | 7 | 22 | 29 | 10 | 14 | 30 | 59 | 299 | 459 |

***Table 1 :** Examples of values given in the first-order case by Wilks formula.*

In Table 1, we present several consistent combinations of the sample size *n*, the quantile order γ and the confidence level β. For example, to have an estimation of the median (0.5-quantile) with a level of confidence of 90%, the Wilks formula requires a sample of size 5 and the corresponding Wilks first-order 0.5-quantile should be the maximum of the sample.

The equation (Eq. 1) is a first order equation because the upper bound is set equal to the maximum value of the sample. To extend Wilks formula to higher orders, we consider the n-sample of the random variable X sorted into the increasing order: $X_{(1)} \leq X_{(2)} \leq \cdots \leq X_{(r)} \leq \cdots \leq X_{(n)}$. For all $1 \leq r \leq n$, we set

$$G(\gamma) = \mathbb{P}[\mathbb{P}(X \leq X_{(r)}) \geq \gamma] \quad \text{(Eq. 3)}.$$

According to the Wilks formula, the previous equation can be recast as

$$G(\gamma) = \sum_{i=n-r+1}^{n} C_n^i \gamma^i (1-\gamma)^{n-i} \quad \text{(Eq. 4)}.$$

The value $X_{(r)}$ is an upper-bound of the γ-quantile with confidence level β if $1 - G(\gamma) \geq \beta$.

Increasing the order of Wilks formula helps reduce the variance in the quantile estimator, the price being the requirement of a larger n (according to formula $1 - G(\gamma) \geq \beta$). This Wilks formula can be used in two ways: when the goal is to determine the sample size required to estimate a γ-quantile with a given confidence level β; when a sample size is already available, then the wilks formula can be used to determine the couple (γ, β) and the order for the estimation of the Wilks quantile.

## 2.2. Application to estimate an upper bound of the $H_2$ production

---

[1] $q_\gamma$ is a $\gamma$-quantile of the random variable $X$, if $P(X \leq q_\gamma) \geq \gamma$, which means $q_\gamma$ is the value for which $\gamma$ proportion of the population is lower.

Our target is to determine an upper bound of the $H_2$ production for the PEGASE facility drums which has been set to a 90% quantile. Therefore, with the Wilks method, the required sample measurement size to estimate a 90% quantile for the studied population of drums with a level of confidence of 90% is 38. Nevertheless, as the measurement cost and operation constraint are important, it has been decided to use as far as possible available measurements of 15 repackaged drums but with the necessity to have a representative sample of the whole population. The sample drum selection is performed according to two quantitative (Pu quantity and the production date) and two qualitative drum parameters (origin and spectrum) considered by expert judgement as the more relevant from the 17 characterising the studied population of drums. Therefore, we use a method consisting in a random selection associated to a relevant parameter representativeness validation with statistical tests (Kolmogorov-Smirnov, Ansari-Bradley) for the numerical characteristics and the use of multinomial and binomial distribution laws to represent the qualitative characteristic occurrence in the whole drum population of PEGASE facility (Pérot 2010). The *origin* parameter/variable is modeled by a multinomial distribution law and the *spectrum* variable is modeled by binomial distribution law. In both case, the law parameters are fitted on the data of the whole drum population. Statistical tests are used to evaluate for each modality of each qualitative variable, if the proportion of occurrence in the random sample is representative of the whole population. If not (for quantitative or qualitative variables), a new random sample is selected.

Thus a sample of 38 drums has are identified including 15 available drum measurements and with only 23 drums to measure. The summary statistics estimated with the H2 flow rate data are the following (the unity is l/year): mean $\hat{\mu} = 2.18$, median = 1.43, standard deviation $\hat{\sigma} = 2.67$, Min = 0.02, Max = 13.97. Figure 1 presents measurement data dispersion on a histogram and a boxplot which seems to be closed to a lognormal distribution in the central part.

We can directly estimate the 95%-quantile from the log-normal theoretical distribution that has been fitted, $\mathcal{LN}(0.23, 1.16)$ :

$$q_{95\%} = 8.5 \, l/drum/year.$$

However, due to the small number of data that served to fit the probabilistic density function, little confidence can be accorded to this value, and justifying it to safety authorities could be difficult. Moreover, the log-normal distribution is rejected by the Shapiro-Wilks adequacy test (the most robust test for small sample size) with the threshold 5%.

A Wilks unilateral second-order $\gamma$-quantile with a confidence level $\beta$, is deduced from equation *(Eq. 4)* and we obtain the following solution for $(\gamma, \beta) = (90\%, 90\%)$:

$$\boldsymbol{q_{\alpha,\beta}^w = q_{90\%,90\%}^w = 8.3 \, l/drum/year}.$$

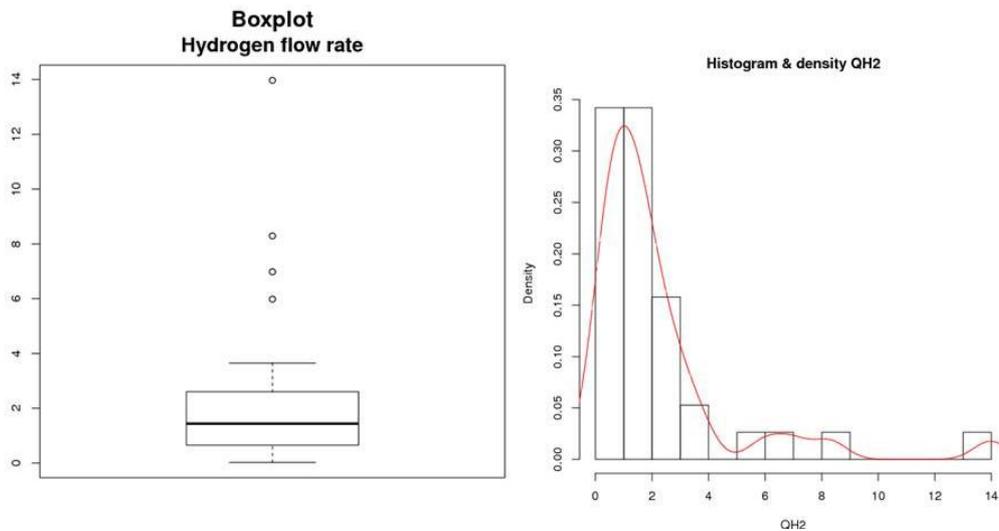

**Figure 1** : Boxplot (left), histogram and smoothed-kernel density function (right) of the 38 hydrogen flow rates.

The statistical analysis of the sample measurement data provides estimations of mean and 90%-quantile for $H_2$ flow rate for an equivalent standard drum (corresponding to 4.5 times the quantity of interest for a primary drum). This estimation is lower than the target threshold of 10 l/year/drum and presents a good convergence using a global measurement from the ventilation system of the facility (Pérot 2010). The sampling strategy (Pérot N. & al 2018), presented in this section is based on the Wilks method to determine the size of the drums to measure and the use of statistical tests to validate the representativeness of the measurement data according to relevant parameters. The aim of the following part is to confirm that the hypothesis used to guide the sampling strategy are valid.

## 3. Validation of sampling strategy hypothesis and regression analysis

### 3.1. Factorial analysis and validation of sampling strategy hypothesis

The sampling strategy performed to estimate upper bounds of $H_2$ flow rate of the drum population is based on a sampling strategy according to 4 parameters considered as relevant. The question is to validate this hypotheses provided by expert judgement. For this purpose, a factorial analysis for mixed data (FAMD) (Saporta 1990; Pagès 2014) is implemented on the 38 data of the measurement sample completed by a hierarchical cluster analysis. FAMD method is a combination of a principal component analysis (PCA) for quantitative data and a multiple correspondence factorial analysis (MCA) for qualitative data.

PCA is a statistical method that use an orthogonal transformation to convert a set of numerical observations of possibly correlated variables into a set of values of linearly uncorrelated variables called principal components. The visualization of the projection of the values of the studied variables on the principal components gives information on the existence of linear multiple correlations on these variables.

MCA method is a data analysis technique that can be considered as a counterpart of PCA for qualitative (categorical) data.

Figure 2 presents the correlation circle for the two first components of the principal component analysis (PCA) on which the drum quantitative parameters have been projected. These two components explain almost 60% of the data variability and the correlation circle highlights. In particular, as expected, we can see a strong multiple correlation for *Nuclear material, Mass of 239Pu, Mass of 239Pu and 241Pu, Mass of 241Pu and Mass of Pu* on the one hand and another one for *Contact dose rate, Dose rate at 1m, Total mas of U and Pu, Mass of depleted U* on the other. Figure 3 presents the results of the analysis for the qualitative variable, the modalities have been projected on the two first components and shows the proximity of the first modality of the variable o*rigin* and the first modality of the variable *spectrum*.

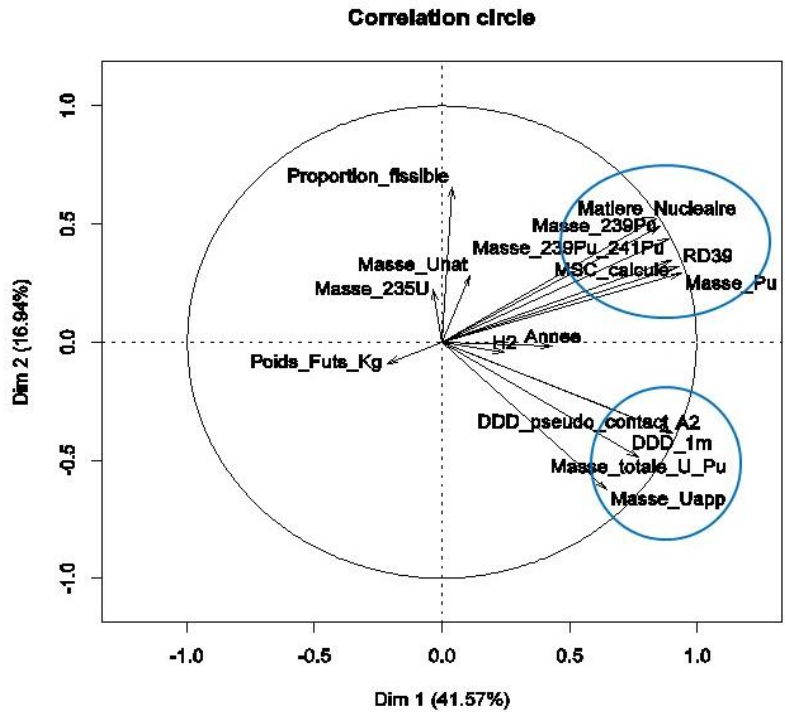

**Figure 2:** Correlation circle of the CPA on quantitative data for the two first components.

A hierarchical cluster analysis is performed on the first component; the results presented on Figure 4 provide 5 clusters with the characteristics detailed in Table 2. For quantitative parameters/variables (resp. qualitative), correlation are the squared correlations (resp. the correlation ratios) with the first principal component (the cluster center). We can observe that the 4 given relevant initial parameters (from expert judgement) are finally cluster representative although *Year of production* and *origin* are in the same cluster. Nevertheless, two clusters are not taken into account; one should be represented by the parameter *Mass of depleted U* and the cluster containing *Mass of Unat* and the specific information parameter *Classification* is not significant considering the correlation values of 0.65. As for the other quantitative criteria, we have performed statistical tests to validate the representativeness of the 38 data selection for the criteria *Mass of depleted U.*

An FAMD performed on the entire population of drums provides the same results and contribute to validate the representativeness of the data selection.

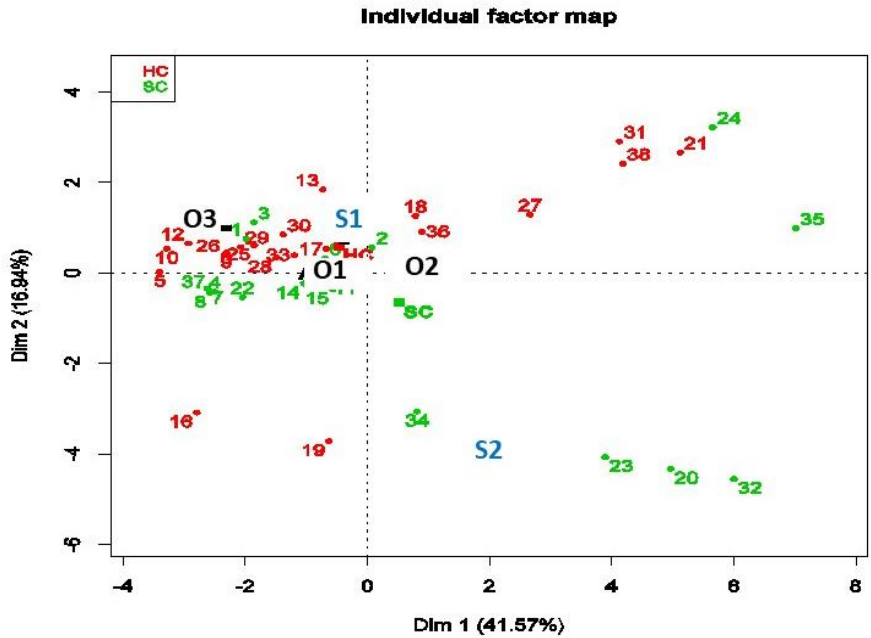

**Figure 3:** Representation of the qualitative variable modalities on the two first components of the FADM analysis.

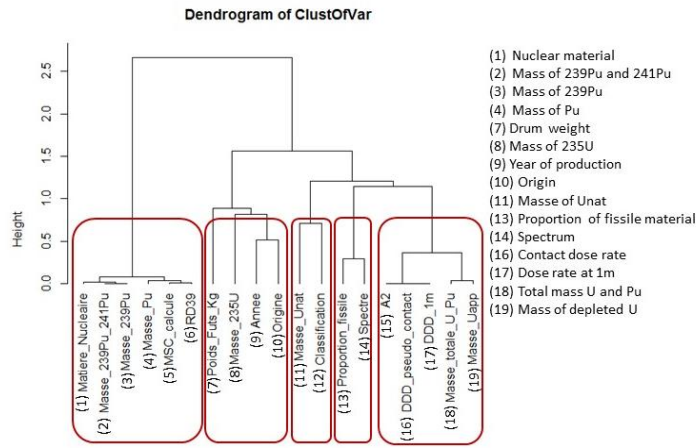

**Figure 4:** Clusters resulting from the hierarchical clustering analysis performed on the sample of 38 measurement data.

| Cluster 1 | | Cluster 2 | | Cluster 3 | | Cluster 4 | | Cluster 5 | |
|---|---|---|---|---|---|---|---|---|---|
| Parameter | Correlation | Parameter | Correlation | Parameter | Correlation | Parameter | Correlation | Parameter | Correlation |
| *Date of production* | 0.67 | *Proportion of fissile material* | 0.86 | Total Mass U and Pu | 0.91 | *Mass of Pu* | 0.97 | Mass of Unat | 0.65 |
| Mass of 235U | 0.31 | *Spectre* | 0.86 | Mass of depleted U | 0.83 | Mass of 239Pu and | 0.98 | Classification | 0.65 |
| *Origin* | 0.58 | | | Contact dose rate | 0.95 | Mass of 239Pu | 0.98 | | |
| Drum weight | 0.23 | | | Dose rate at 1m | 0.95 | Nuclear material | 0.98 | | |

**Table 2:** Description of the clusters provided by the hierarchical cluster analysis.

### 3.2. Regression analysis

Both qualitative parameters "origin" (3 modalities) and "spectrum" (2 modalities) used for the sampling strategy and the clustering analysis provides 5 main clusters (some combination are not realistic), and for each of them, it could be developed a prediction model for the $H_2$ flow rate production. Nevertheless, the size of these clusters is very different in the measurement sample. Indeed, there are 22 data for the cluster corresponding to the origin "O1" and the spectrum "S1" but there are only 7 data for the cluster corresponding to the origin "O2" and the spectrum "S1". The other clusters are even smaller. For a robust and reliable regression analysis, the studied sample has to be of sufficient size.

A regression analysis is performed on the cluster "O1-S1" (which corresponds to the closest modalities highlighted by the FADM analysis in the previous section) with 22 data. A general linear model regression with the elastic net regularization method (Zou & al. 20015) is performed that combines the $L_1$ and $L_2$ penalties of the Lasso and Ridge methods to reduce the number of terms in the regression model and identify the important terms or predictors. This method treats the redundant terms and produces estimates with potentially lower predictive errors than ordinary least squares. The study used *glmnet R* package.

Therefore, performing this regression method, we obtain a linear model with 4 parameters (*Production date*, *Mass of depleted U*, *Mass of 235U* and *Mass of Pu*) which represent the 4 main clusters provided by the hierarchical cluster analysis:

$$Q_{H_2} = -4.73 + 0.07 \times Production_{date} + 0.026 \times Mass_{of\,Pu}$$
$$+0.006 \times Mass\_of\_depleted\_U \times Mass\_of\_235U \qquad (Eq.\ 5)$$

The determination coefficient $R^2 = 75\,\%$, and the adjusted determination coefficient (which take into account the number of data and the number of model terms) $R_a^2 = 67\,\%$ are not close to 1 and do not help to validate the model. Moreover, the quantile-quantile graph of Figure 5 which represents the plot of the theoretical model quantiles against the data empirical quantiles does not show a good adequacy of the model *(Eq. 5)* for a high quantile represented by the 37[th] data that could be an outlier or a misinformed data. In order to improve the model of $H_2$ production rate prediction, the regression is performed on the data of cluster "O1-S1" without the 37[th] data.

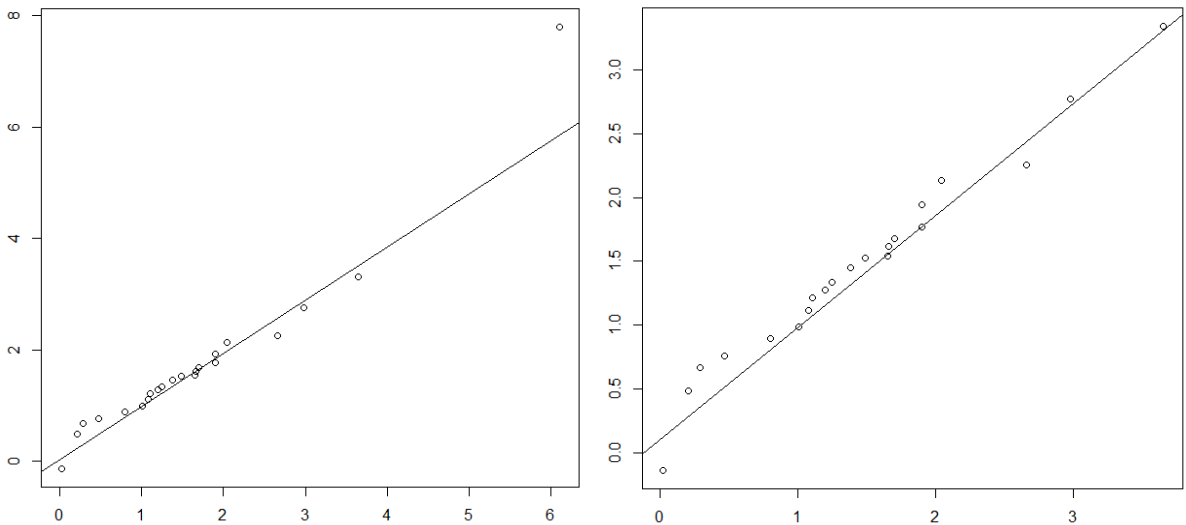

**Figure 5:** Quantile-quantile graph of the model *(Eq. 5)* versus the measurement data for the cluster *"O1-S1"* (global on the left and without the higher quantile on the right).

We obtain the following linear model:

$$Q_{H_2} = -5.52 + 0.078 \times Production_{date} + 0{,}014 \times Mass_{of_{Pu}} \\ + 0{,}0058 \times Mass\_of\_depleted\_U \times Mass\_of\_235U \qquad (Eq.\ 6)$$

The determination coefficient $R^2 = 81\,\%$, and the adjusted determination coefficient $R_a^2 = 78\,\%$ for the model *(Eq. 6)* are closer to 1 than for the model (Eq. 5) and contribute to validate this model. Moreover, the quantile-quantile graph of Figure 6 shows a better adequacy for the central quantiles than for the previous model.

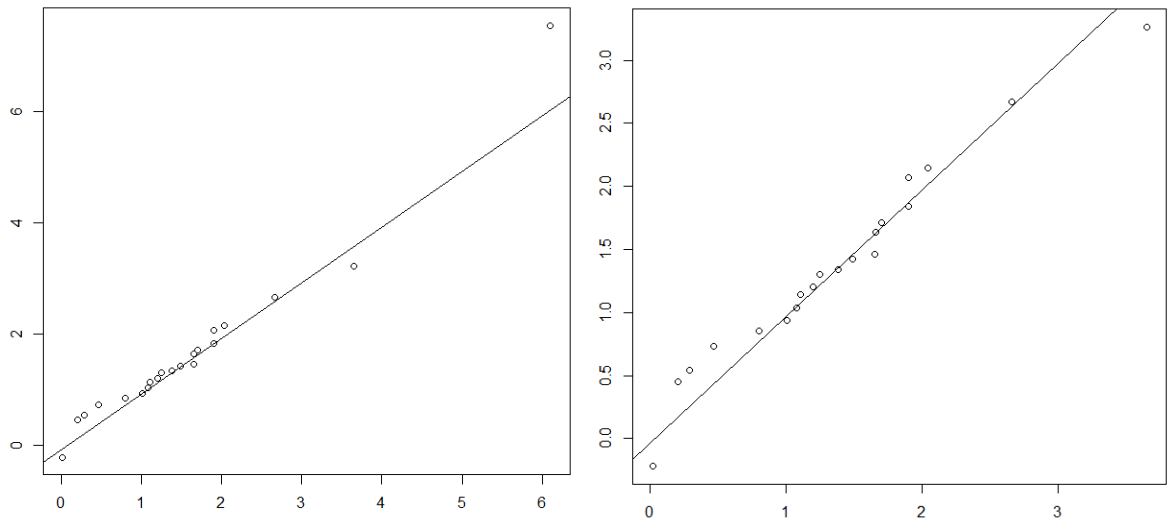

**Figure 6:** Quantile-quantile graph of the model *(Eq. 6)* versus the measurement data for the cluster *"O1-S1" without* the 37[th] data (global on the left and without the higher quantile on the right).

This model *(Eq. 6)* is then used to provide confidence intervals of $H_2$ flow rate production for the 1462 drums of the studied cluster "O1-S1" in the whole population and the following confidence intervals of level 95% and 90% are estimated:

$I_c^{95\%} = [4.9;\ 12.9]\ l.\,year^{-1}$ for the drum with the higher predicted $H_2$ production rate,
$I_c^{90\%} = [5.6;\ 12.2]\ l.\,year^{-1}$ for the drum with the higher predicted $H_2$ production rate.
The estimation of the predicted $H_2$ production rate for this drum is $8.88\ l.\,year^{-1}$.

## 4. Conclusion

We have presented here a methodology of sampling strategy to select a small sample to be measured in order to provide estimations of the mean and the upper bound of a physical quantity of interest. The results of the statistical analysis give a good convergence with global measurements from the ventilation system of the facility. Thereafter, performing a factorial data analysis demonstrates and completes the sampling strategy assumption validity (assumptions derived from expert judgement) and provides information that are used for a regression analysis to develop a linear prediction model of dihydrogen flow rate production for the waste drum characterization of the studied facility.

This approach has demonstrated its value, so it is currently performed in order to characterize other physical quantities for waste drums on several CEA facilities.

## Acknowledgment

The authors wish to thank Thierry Advocat and Eric Fillion, from CEA D&D research program for their support.